\documentclass[twocolumn,letterpaper,superscriptaddress]{revtex4}
\usepackage{graphicx}
\usepackage[usenames]{color}
\usepackage{amsmath}

\begin{document}
\title{Graphene nanodevices: bridging nanoelectronics and 
subwavelength optics  }
\author{Pierre \surname{Darancet}}
\affiliation{Institut N\'eel, CNRS \& UJF, Grenoble, France}
\affiliation{European Theoretical Spectroscopy Facility (ETSF), France}
\author{Valerio \surname{Olevano}}
\affiliation{Institut N\'eel, CNRS \& UJF, Grenoble, France}
\affiliation{European Theoretical Spectroscopy Facility (ETSF), France}
\author{Didier \surname{Mayou}}
\email[corresponding author: ]{didier.mayou@grenoble.cnrs.fr}
\affiliation{Institut N\'eel, CNRS \& UJF, Grenoble, France}
\affiliation{European Theoretical Spectroscopy Facility (ETSF), France}

\date{\today}

 \begin{abstract}
The unconventional properties of graphene, with a massless Dirac band
dispersion and large coherence properties, have
raised a large interest for applications in nanoelectronics.
In this work, we emphasize that graphene two dimensional character
combined with current standard lithography processes allow  to achieve devices
smaller than the Dirac electrons wavelength.  In this regime, we demonstrate
that the electronic properties present deep analogies with subwavelength
optics phenomena. We describe the rich transport physics in
graphene-based nanodevices through optical analogies: From the Bethe 
and Kirchhoff-like diffraction patterns in
the conductance of graphene slits to the Fabry-Perot oscillations of
the conductance in nanoribbons. We introduce the  concept of {\it
electronic diffraction barriers}, which transmission cancels at the 
Dirac point. This gives  central insight in
the properties of Graphene subwavelength devices including 
nanoelectronics standard systems, such as quantum dots.
As an application we propose a new type of quantum dots,
namely functionalized subwavelength quantum dots, which could be used 
as  molecular spin valves.
\end{abstract}

\maketitle

%
%

Analogies play a prominent role in physics.
By allowing transfer of notions and advances from one field to another, 
they can provide deeper insight in those fields.
In this context, the analogy between  optics 
and quantum electronic transport, thanks to the ondulatory nature of 
electrons and light, has been particularly useful either in predicting the 
transport properties of devices  \cite{G}, or in understanding optical phenomena such as the coherent 
multiple scattering of light \cite{Wolf}.
Indeed, as long as the coherence lengths of a material are large, 
the -quantum transport- Landauer formula establishes a one to one correspondence
 between conductance and transmission in scattering experiments.
While light scattering  experiments probe the transmission probabilities between a given 
pair of input and output modes, the transport experiments probe the sum of the transmission 
probabilities from all possible input modes to all possible output modes at the Fermi energy, 
\textit{i.e.} at a given frequency in the optical language. 

Such analogies are particularly relevant in  graphene -and have already been exploited \cite{Altshuler}-
due to extraordinary coherence properties of this material. Indeed, this system offers unique characteristics 
either in the exfoliated or in the epitaxial form 
\cite{Novoselov,Kim,Berger,Berger2}. The electronic mean-free path can be of 
the order of a micron at room temperature and the Dirac electrons
wavelength, which in principle diverges at the Dirac point,  can be 
up to 100 nm or even more \cite{A}, in real systems. Moreover,
graphene is at the surface and directly accessible to
lithography, allowing to  produce nanodevices down to a few  nanometers  as typical
size \cite{B,B2}, \textit{i.e.} smaller than the Dirac electron wavelength.

In this work, we demonstrate that the transport properties of such devices can be understood
by bridging nanoelectronics of 
graphene nanodevices with the domain of subwavelength optics.
Indeed, at this scale, quantum interference effects in graphene become prominent, and 
the conductance exhibits a universal behaviour characteristic of 
subwavelength optics.
We believe that this analogy between two fields,  where intense effort are made to find devices with original properties, can be very fruitful and we give  examples in this work. 

We consider first systems  such as  slits and nanoribbons  sandwiched in between graphene semi-infinite sheets
as leads, which have well known optical analogues. These studies lead us to the 
mixed "electronics-optics" concept  of  \textit{electronic diffraction barriers} which transmission cancels for a divergent wavelength of the electron in the graphene plane, \textit{i.e} at the Dirac point.
 This concept  gives also an important insight in the confinement in graphene quantum dots where a
chaotic Dirac billiard behaviour has been recently observed \cite{Geim}. 
It shows that the characteristics of these  quantum
dots, such as barrier transparencies, can be
tuned and controlled, thus opening to a wealth of new potential
applications. A new type of quantum dot is proposed by
functionalizing carbon atoms in a graphene constriction. In
particular functionalization by magnetic molecules could provide molecular spin valves.  
 
\begin{figure}
      \includegraphics[clip,width=0.45\textwidth]{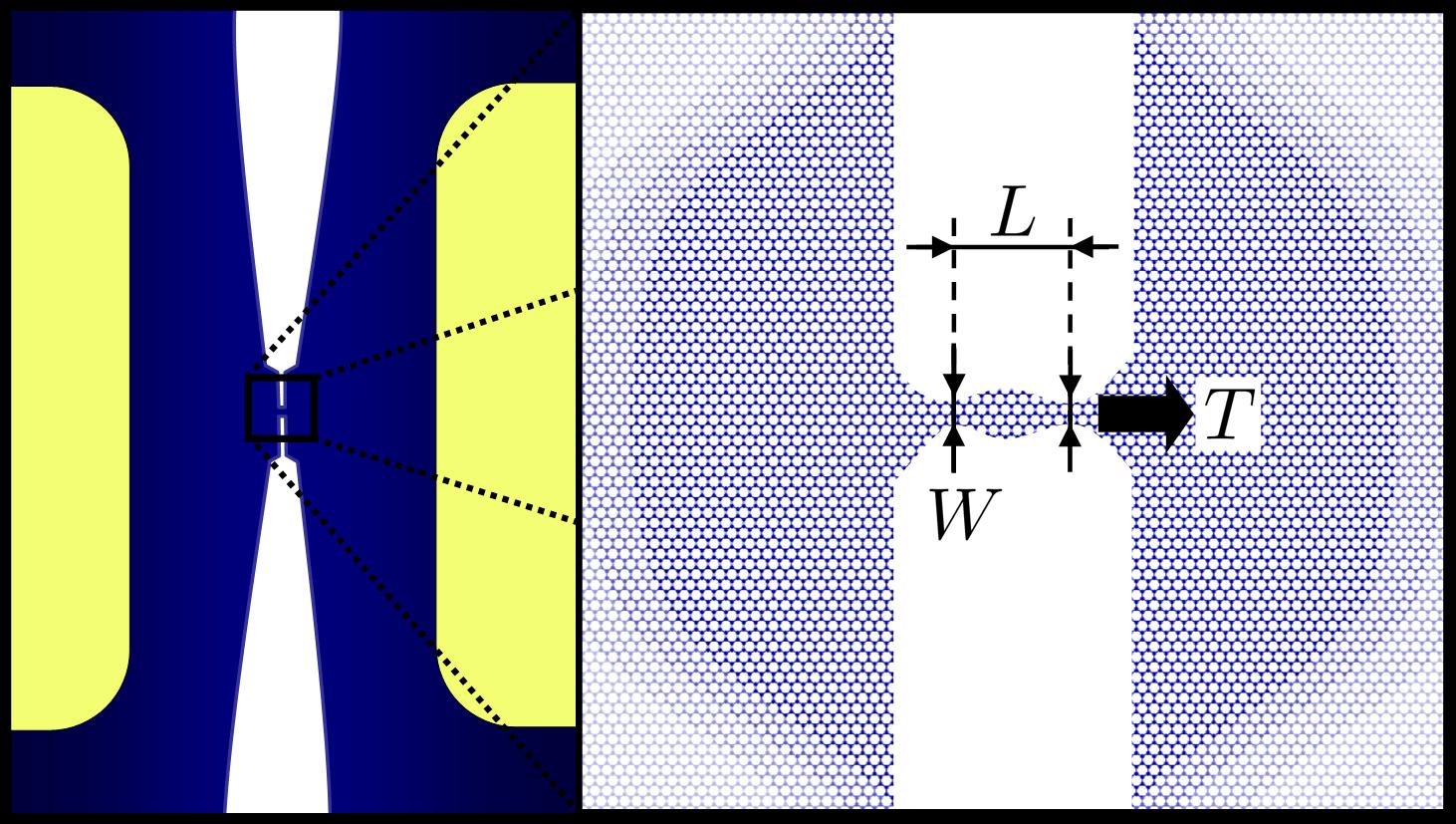}
      \caption{\textit{Schematics of the systems studied in this article.} Left:
the yellow part represents the metallic contacts and the blue region
is the graphene part of the device. Right: detailed view of the
tripartitioned system. The typical width of the contact is $W$ and
the typical length of the central device is $L$.}
      \label{general}
\end{figure}

The prototypical system we consider is  shown in Fig.~\ref{general}.
It is a tripartitioned system consisting of  a central device (a
graphene nanoribbon, a slit or a graphene quantum dot), contacted to
mesoscopic leads made of  two semi-infinite graphene half planes. The
system parameters which affect the response of the system with
respect to optical-like phenomena are the aperture $W$ of the
contact, the typical length $L$, and the energy dependent electron
wavelength $\lambda$. The chirality and the electronic structure of
the ribbon can also affect the conductance of the device. Yet we will
show that the quantum interference effects studied here present a
universal  behaviour, independent of the electronic
structure of the system, that present a deep analogy with phenomena 
that are known in subwavelength optics \cite{Ozbay,Barnes}.

Our methodology relies on an exact numerical calculation of the
conductance within a tight-binding \cite{C} and a Landauer quantum
transport formalisms. The novelty  is that the Landauer
equation is solved via a new recursive numerical algorithm which
reduces complex 3D/2D various shaped quantum transport devices into
an effective 1D system. Here only the channels
effectively contributing to the conductance are considered; the
method  thus allows to exactly compute the contact resistance in an
efficient way.

\begin{figure}
      \includegraphics[clip,width=0.45\textwidth]{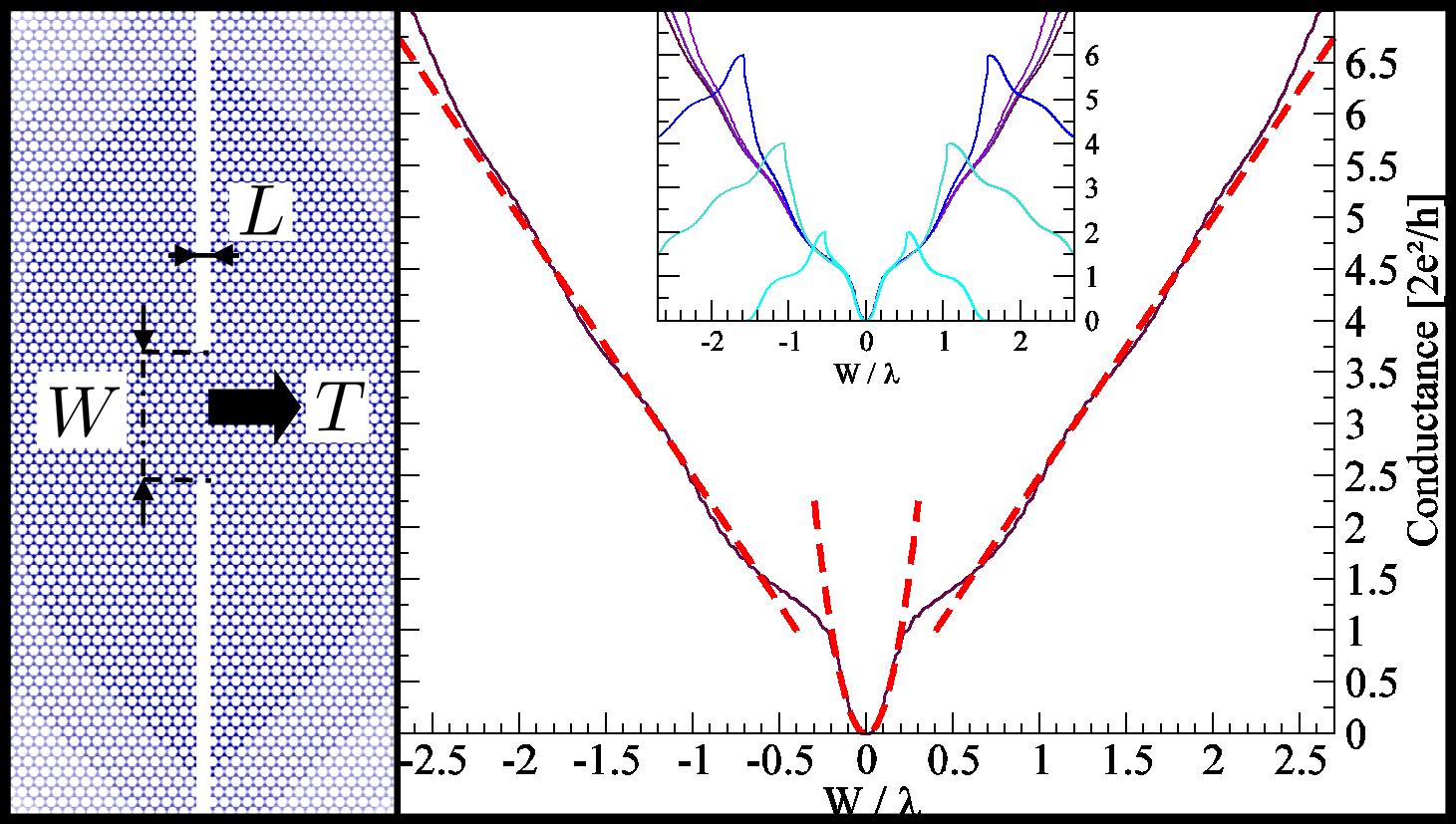}
      \caption{\textit{Conductance of  graphene slits and diffraction} Left:
schematic view of the geometry of a graphene slit. Right: Scaling law for
a graphene slit conductance  as a function of $W/\lambda$. This shows
the Bethe-like $\lambda \gg W$ quadratic regime and the
Kirchhoff-like $\lambda \ll W$ linear regime (see text). The inset
shows the conductance of slits with different widths.}
     \label{diffraction}
\end{figure}

\textit{Conductance of  graphene slits and diffraction} -
The archetype experiment  showing an optical behaviour is the 
transmission through a hole, for example in the geometry indicated in the left panel of Fig. \ref{diffraction}.
The conductance of the slit is related to the transmission of eigenstates (or modes) 
through the slit. Therefoore  the situation is obviously connected to the 
textbook experiment of diffraction by a hole. In optics depending on 
the hole width relative to the wavelength   there are 
different regimes. The standard Kirchoff theory applies only in the 
large wavelength limit. However the phenomenon of diffraction through 
a subwavelength slit or through an array of such slits is central to 
subwavelength optics and plasmonics \cite{Ozbay,Barnes,Ebbesen}. 
Analogeous results are detailed here forthe conductance of graphene 
slits.

Here we report the conductance of several slits that
consist of  a single \textit{motif}, {\it i.e.} a single hexagon  (the
edges of the graphene half-planes being of the armchair type) but 
differing by the width $W$
(Fig.~\ref{diffraction} inset). We introduce the wavelength 
$\lambda$ of the incident Dirac
electrons:

\begin{equation}
     \lambda = \frac{2\pi}{k} = \frac{h}{p} = \frac{h v_{\rm F}}{E}
     \label{lambda},
\end{equation}
where $k$ is the Dirac electrons wavevector, $p$ is the momentum, $E
= v_{\rm F} p$ is the energy, linearly dispersed in graphene with a
Fermi velocity of $v_{\rm F} \simeq 1 \cdot 10^{6}$ ms$^{-1}$
\cite{D}.
 
 We first note (Fig.~\ref{diffraction} inset) that for all the
slit widths $W$ a universal scaling occurs such that the conductance 
depends only on the ratio $W/\lambda$. This scaling breaks down when
the wavelength is $\lambda \lesssim 4$ nm, that is the energy is
greater than approximately $1$ eV. Indeed as shown below this scaling 
is intimately related to the scaling properties of the Dirac equation 
which  is valid only in the low energy or long wavelength limit.

We then note that in the limit $W/\lambda \ll 1$, the
conductance is quadratic, with the conductance $g(W/\lambda)\simeq 25
(W/\lambda)^{2}$, while in the opposite limit $W/\lambda \gg 1$, the
conductance turns out to be linear with  $g(W/\lambda)\simeq 2.5
(W/\lambda)$. The crossover between the two regimes occurs around
$W/\lambda\simeq 0.3$. This promptly reminds of the analogy with
classical optics and hence offers an immediate interpretation. We can identify two different
diffraction regimes: for wavelengths $\lambda \gg W$, much larger
than the aperture, we observe a {\it Bethe-like diffraction regime}
\cite{Bethe} with a slit transmitted energy proportional to the
square of the aperture. On the other hand, for $\lambda \ll W$ one
has a {\it Kirchhoff-like diffraction regime}, the transmitted energy
being proportional to the aperture. Thus the fact that the system
does not conduct at the Dirac energy is a manifestation not of the
system electronic structure, but rather of the classical optical
inability of an incident wave to be transmitted through a
subwavelength constriction which hence behaves like a barrier.
 
The explanation of both the observed diffraction patterns and of the
universal scaling law comes from  analogies between Maxwell
equations and the Dirac equation here. Indeed, a universal scaling is
also observed in ordinary optical diffraction for the slit
transmitted intensity due to the scaling invariance of Maxwell
equations. Here it stems from the fundamental scaling invariance of
the graphene 2D Dirac equation which can be written in a
dimensionless form,
\begin{equation}
     \hbar v_{\rm F} \mbox{\boldmath$\sigma$} \mbox{\boldmath$\cdot$}
\frac{ \partial \psi}{\partial \mathbf{r}} =  E \psi
     \quad \Longrightarrow \quad
      \mbox{\boldmath$\sigma$} \mbox{\boldmath$\cdot$} \frac{ \partial
\psi}{\partial \mathbf{\tilde{r}}} =  2 \pi \psi
      ,
     \label{Dirac2}
\end{equation}
where \mbox{\boldmath$\sigma$} are the Pauli matrices, $\psi$ is a
two component pseudo-spinor, and
$\mathbf{\tilde{r}} = \mathbf{r} / \lambda$. The scale invariant
($\mathbf{r} / \lambda$ dependent) form of the Dirac equation
directly stems from the linear dispersion relation $E=hv_{\rm
F}/\lambda$. This leads to the fact  that
the conductance undergoes a scale invariance $g(x\lambda, xW) =
g(\lambda,W)$, so it only depends from a reduced argument
$W/\lambda$, $g(\lambda,W)=g(W/\lambda)$.

The argument is as
follows. Let us consider a system where each half plane of graphene
is replaced
by a ribbon of width $\tilde{W}$. The  symmetry axis of the ribbon is
perpendicular to the slit, passing by the center of the slit. The
conductance for half graphene plane is
$g(\lambda,W)=lim_{\tilde{W}\rightarrow
\infty}G(\lambda,\tilde{W},W)$. We assume that the boundary conditions
preserve the scaling, which is evidenced here by the numerical results.
This is also the case for continuous type models of confinement
\cite{Footnote2}. Then there is a one to one correspondence between
the scattering states at wavelength  $\lambda$ ribbon width $\tilde{W}$,
slit width $W$  and those at wavelength  $x\lambda$, ribbon width
$x\tilde{W}$  and slit width $xW$. This means that the conductance
satisfies $G(\lambda,\tilde{W},W)=G(x\lambda,x\tilde{W},xW)$ and as a
consequence
$g(x\lambda, xW) =
g(\lambda,W) = g(W/\lambda)$.

\textit{Nanoribbons as subwavelength waveguides} -
Another very 
well-known evidence for interference effects, are the
Fabry-Perot oscillations, which occur when an optical waveguide is 
sandwiched in between two reflecting surfaces.
Several studies of subwavelength
optical  waveguides have shown the existence of Fabry-Perot
oscillations of the transmitted intensity for an incident wave
parallel to the axis of waveguide. This is for instance the case for   strips 
in subwavelength optical metallic gratings \cite{Cavity2} or
subwavelength slits in metallic screens with a finite thickness
\cite{Takakura,Suckling}. In these systems the reflection at the ends 
of the waveguide is in fact intimately related to the diffraction by 
the subwavelength aperture.

In our case, the corresponding $2D$ transport configuration is a 
metallic entity
(corresponding to the optical waveguide), that we chose to be a 
metallic graphene nanoribbon, connected to two semi-infinite planes 
of graphene. Here also as discussed below the diffraction
by the small apertures of the ribbon produces the reflection 
phenomena at the ends of the nanoribbons. Graphene nanoribbons can be realized by standard lithography
techniques \cite{B,B2,J}. We considered a zigzag and an armchair ribbon,
both chosen with a metallic character and hence an available
conductance channel at the Dirac energy \cite{dresselhaus}, which can 
thus be considered from an optics
point of view as waveguides.

\begin{figure}
      \includegraphics[clip,width=0.45\textwidth]{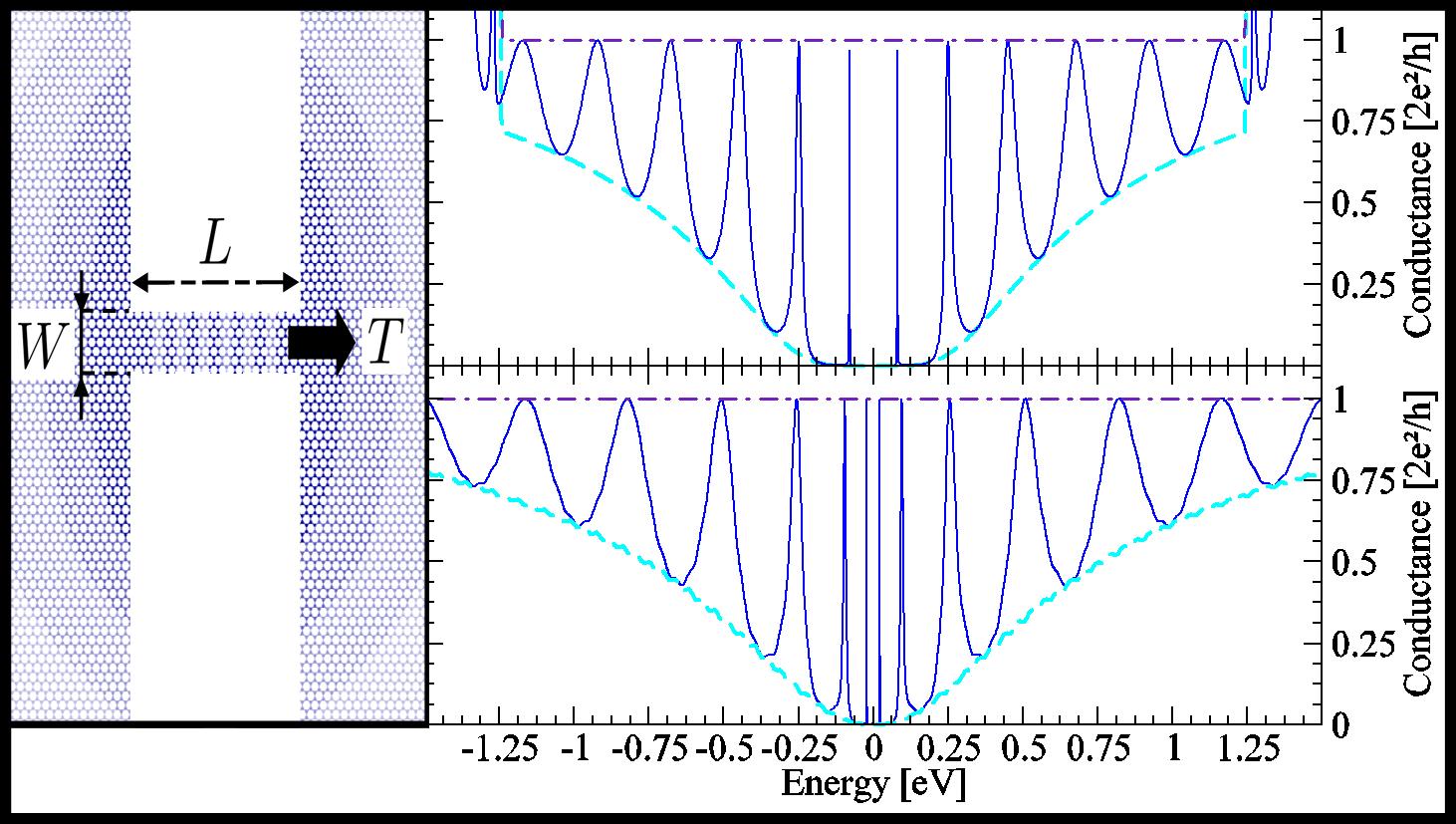}
      \caption{ \textit{Graphene nanoribbons as subwavelength waveguides}
Left: schematic view of nanoribbons contacted with graphene
half-planes. Right: Fabry-Perot oscillations of the conductance of
armchair nanoribbon with a length of $6$ nm (upper part) and zigzag 
nanoribbon with a length of $3$ nm (lower part).  The
envelope of the minima of the conductance oscillations deduced from
Eq.~(\ref{minimaenvelope}) is represented by the dashed line.}
      \label{zigzag}
\end{figure}

Calculating the conductance for both metallic nanoribbons (Fig.~\ref{zigzag}) 
clearly reveals a Fabry-Perot
behaviour. Let us note that these
oscillations exist for perfect nanoribbons but could be
destroyed for example by edge defects \cite{J,K1,K2}. The most 
evident features are
large amplitude {\it oscillations} in the conductance, from maxima values of 1
to minima placed along an envelope function. These are standard 
Fabry-Perot oscillations of  the  transmittance which follow the Airy 
function

\begin{equation}
      T_{\rm FP}(E) = \frac{1}{1 + F(E) \sin^2(\phi(E)/2)}
     \label{T}
\end{equation}
where $\phi(E)= 2 k(E)L + 2 \tilde{\phi}(E)$ is the phase difference
after one loop in the nanoribbon. $k(E)$ is the wavevector of the
Bloch state in the infinite ribbon with energy $E$. $\tilde{\phi}(E)$
is the phase factor acquired at each reflection. $L$ is the length of
the Fabry-Perot interferometer.  $F(E) = 4R(E)/(1-R(E))^2$ is the
{\it finesse} coefficient where $R(E)$ is the reflection coefficient 
at each end of the nanoribbon. 

The Airy function presents maxima $T_{\rm FP}^{max} = 1$ equal to 1
when $\phi/2$ is an integer $m$ multiple of $\pi$.  For sufficiently
large $L$ the phase $\phi(E)= 2 k(E) L + 2 \tilde{\phi}(E)$ varies
rapidly with energy, as compared to $F(E)$; and the minima occur
essentially when $\sin^2(\phi/2)$ is maximum, that is at $\phi/2=m\pi +
\pi/2$, and envelope along the function $T_{\rm FP}^{min}(E)$ which 
independent of the size as we have numerically checked.

\begin{equation}
      T_{\rm FP}^{min}(E) = \frac{1}{1+F(E)} = \frac{(1-R(E))^2}{(1+R(E))^2}.
      \label{minimaenvelope}
\end{equation}

The minimum of conductance tends to zero at small energies 
indicating that $R(E)$ tends to $1$ close to the Dirac energy. 

The peaks full width at half maximum is $\delta \phi_{\rm FWHM} =
2(1-R)/\sqrt{R}$ such as the peaks look thin when $R$ is close to 1,
and broaden when $R \rightarrow 0$. In our case when $E\rightarrow 0$
the peaks seem to have negligible width, indicating again that
$R\rightarrow 1$ close to the Dirac energy. The peaks  broaden for 
the higher energies, indicating
lower values of $R$. 

\textit{Concept of electronic diffraction barrier} -
The previous study on nanoribbons shows the existence 
of Fabry-Perot oscillations and thus a phenomenon of reflexion at 
each end of the ribbon characterized by a reflexion coefficient 
$R(E)$. Just as for the optical analogues this relexion phenomena is 
intimately connected to diffraction at the ends of the ribbon. 

\begin{figure}
      \includegraphics[clip,width=0.45\textwidth]{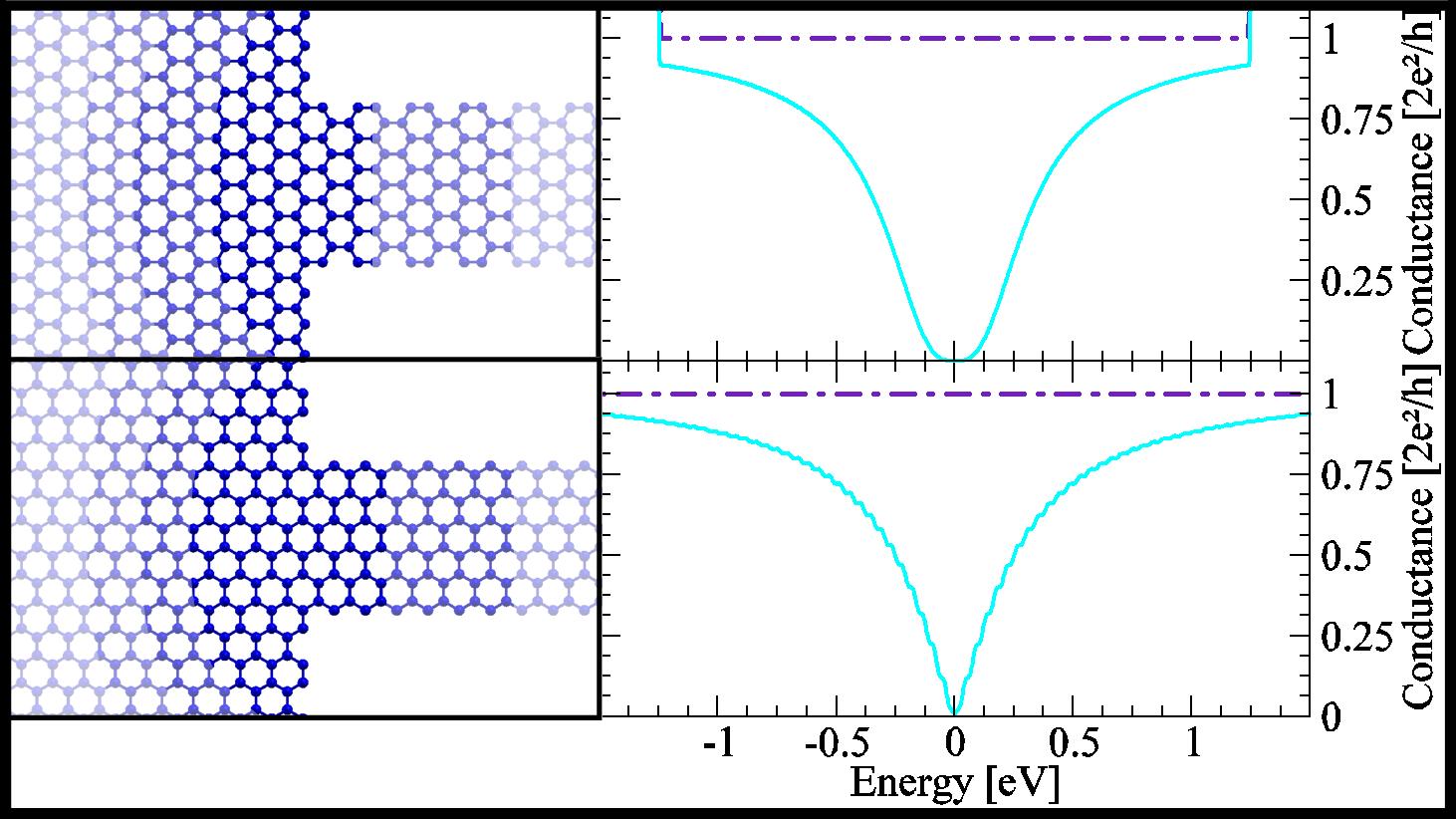}
      \caption{\textit{  Electronic diffraction barrier} at a contact. Left: the
contact  geometries of the armchair (upper panel)  and zigzag (lower
panel) semi-infinite ribbons  coupled to graphene half-planes. Right:
Conductance $g$ of the device represented on the corresponding  left
part. $g=T 2e^2/h$ with $T$ the transmittance of the diffraction
barrier.
}
      \label{armchair}
\end{figure}

The 
reflexion coefficient $R(E)$ is related to the conductance $G$ of  a 
junction between a semi-infinite metallic ribbon and a graphene half-plane
(Fig.~\ref{armchair}) by $G=(2e^2 /h)\times (1-R(E))$. If there was no 
reflexion  we would
expect to observe a conductance equal to 1 quantum of conductance
($2e^2 /h$) (dot-dashed line in Fig.~\ref{armchair}) in the range of
the Dirac/Fermi energy. Instead, the calculated transmittance $T(E)=1-R(E)$
through the junction (continous line in Fig.~\ref{armchair}) is
observed to drop to zero at the Dirac energy, no matter the
chirality, exactly like in slits. This means that \textit{the 
reflexion coefficient $R(E)$ tends to $1$ at the Dirac energy} as 
expected from the analysis of Fabry-Perot oscillations (see above). We 
have checked also that formula \ref{minimaenvelope} gives the correct 
minima of the oscillations with $R(E)$ calculated from the junction 
transmission $T(E)=1-R(E)$. 

Let us stress that $T \neq 1$ behaviour is genuinely a manifestation of the 
contact resistance at the junction, which is independent
of the length of the ribbon, due to its metallic character.
At the opposite, in the case of non-metallic nanoribbons where confinement 
effects create a gap in electronic structure  \cite{Berger,Berger2}, the average resistance increases
exponentially with the length of the ribbon.

The junction conductance we
have calculated here is an example of the elementary
{\it electronic diffraction barrier} which is at the basis of the
universal behaviour observed in the devices hereby studied.

\textit{Application to quantum dots} -
With the  concept of \textit{electronic diffraction barrier 
at a contact} we can now  analyse the conductance of 
standard systems of nanoelectronics \textit{i.e.} quantum dots. We 
consider two examples that illustrate the role of geometry and the 
role of a chemical functionalisation of the dot . 
 
\begin{figure}
      \includegraphics[clip,width=0.45\textwidth]{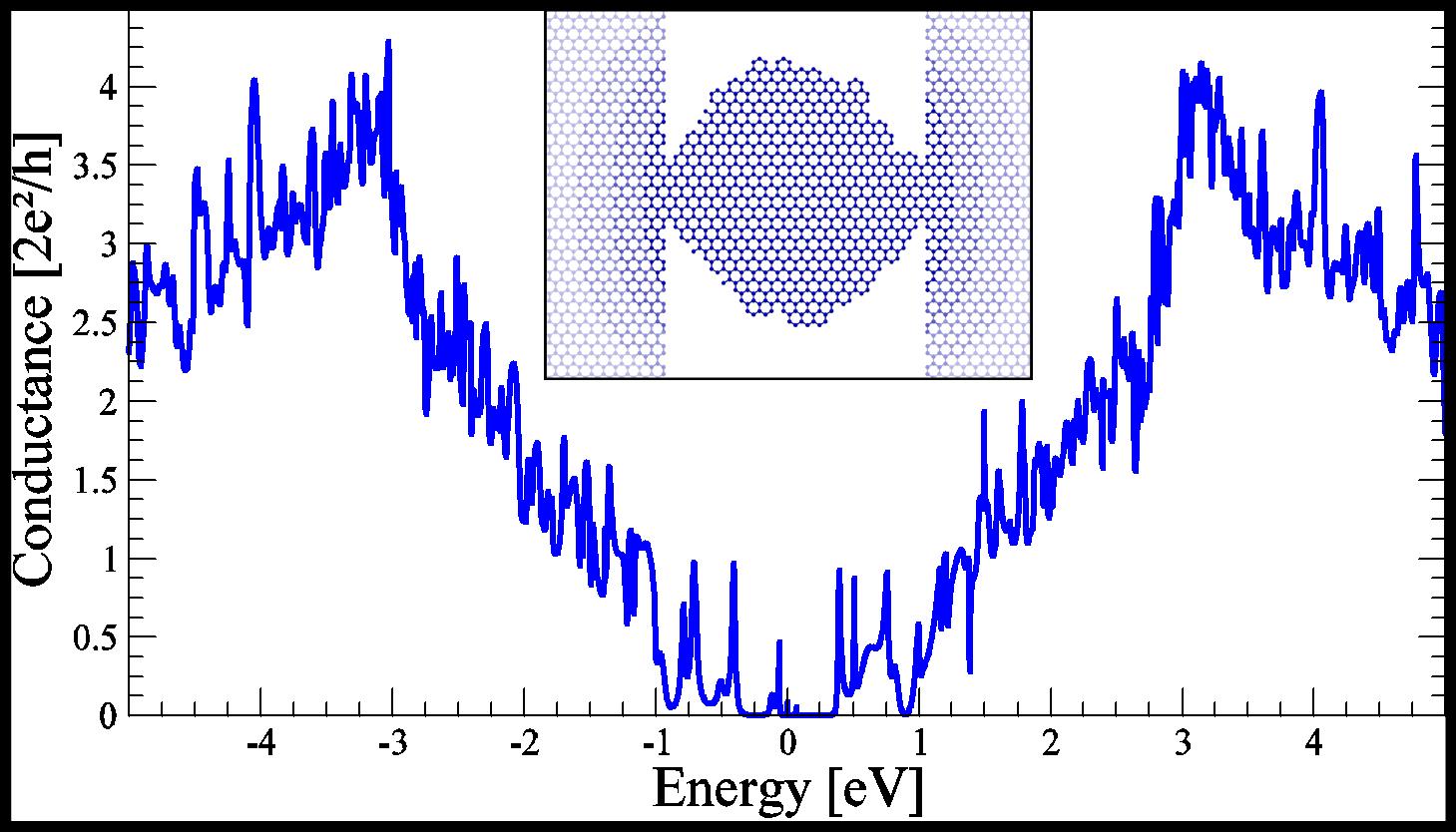}
      \caption{\textit{Effect of the geometry on  the conductance of 
quantum dots}  showing well defined
resonances close to the Dirac energy.  Inset: geometry of the
irregular shaped quantum dot coupled to the graphene half-planes. }
      \label{quantumdot1}
\end{figure}

We consider 
first  a graphene
quantum dot (Fig.~\ref{quantumdot1}) consisting of  a purposely
irregular shape graphene nanostructure contacted via small apertures
to the two half-planes graphene leads. From the concept of electronic 
diffraction barrier we expect that the quantum dot
is weakly coupled to the leads when the energy of the electrons in 
the graphene sheet
is close to the Dirac energy, {\it i.e.} when their wavelength is
sufficiently large.

We note that  the conductance of this system (Fig.~\ref{quantumdot1})
presents a maximum around $3$ eV  which is  the energy of the
Van-Hove singularity of bulk graphene where the density of states
diverges. We also note many irregular peaks superimposed to the
envelope, clearly showing characteristic resonances of the quantum
dot to be associated to the particular (irregular) shape. The
conductance decreases when approaching to the Dirac point, which is a
signature of the lowest number of available states and
of a diffraction barrier at the contact constrictions.

The maximum to minimum transmission ratio  between adjacent peaks is
much larger in the vicinity of the Dirac energy than at other
energies, 
a phenomena which  is also observed in
the nanoribbons studied previously. This is due to the two electronic 
diffraction
barriers at the contacts which are nearly completely reflecting 
($R=1$) at the Dirac energy. This leads
to well defined states (or modes) within the quantum dot, and thus to
transmission channels  at well defined energies  through these states.

\begin{figure}
      \includegraphics[clip,width=0.45\textwidth]{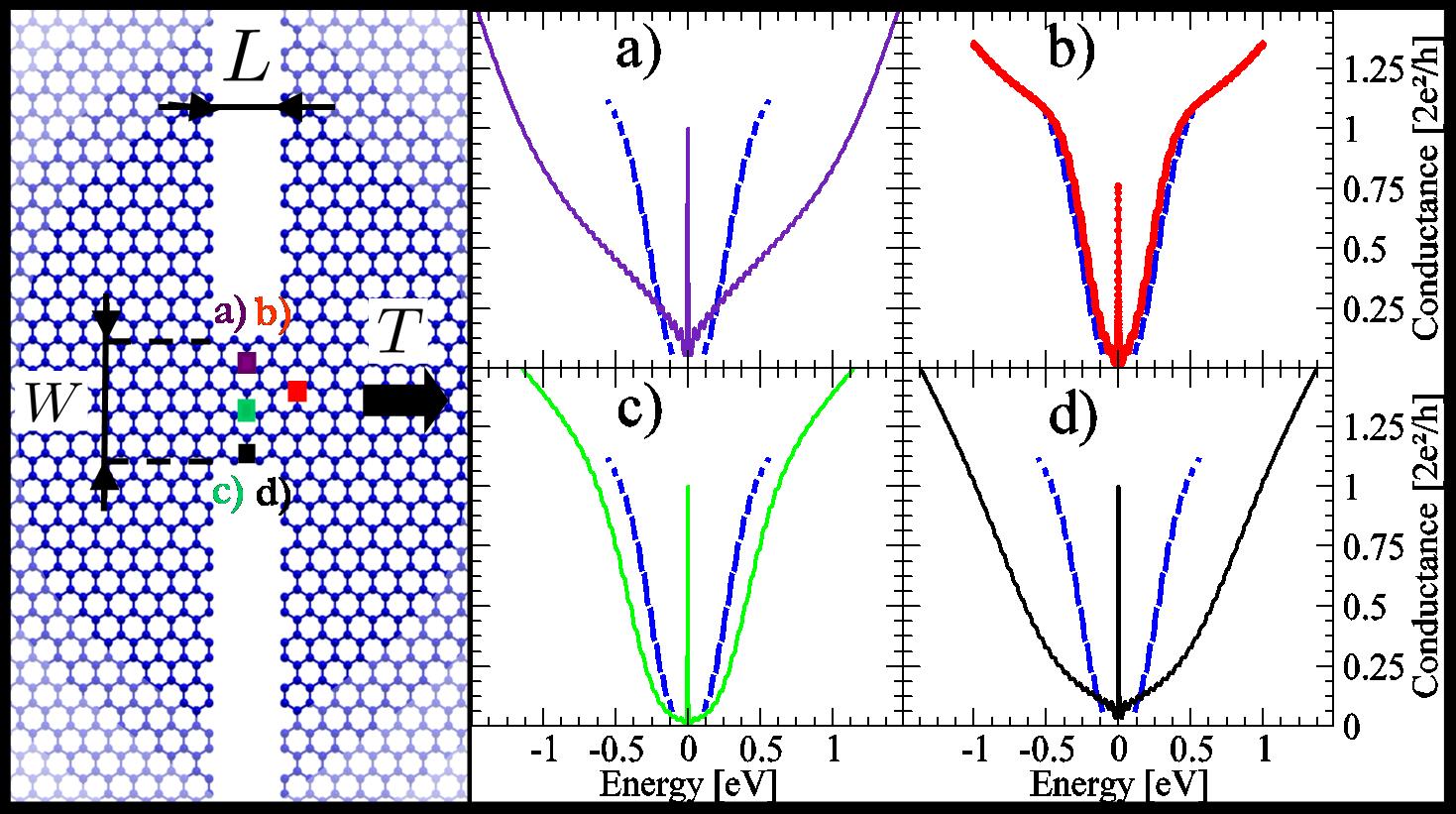}
      \caption{\textit{Effect of a chemical functionalization on  the 
conductance of quantum dots}
Left: geometric configuration with the different postions of the
functionnalized atom. Right: conductance curves corresponding to the
different positions of the functionnalized atom. The blue curve is
the reference  conductance {\it i.e.}  before functionalization.}
      \label{quantumdot}
\end{figure}

A second example  is the effect of the
functionalization by impurities on the response of a graphene quantum
dot. Here our quantum dot is a very short graphene nanoribbon
contacted by small apertures to half-plane graphene leads. This is
like in the slit geometry studied above. The novelty is that now
resonances in the quantum dot are induced by functionalization.
A carbon atom is removed at a given
position and this should mimic the effect of functionalisation by chemical
entities like OH or other groups that strongly hybridize with the
carbon $p_{z}$ orbitals \cite{E}. Due to the specific electronic
structure of graphene, functionalization of one carbon atom in the
plane  creates a so-called vacancy state at the zero ({\it i.e.} Dirac)
energy \cite{F}. This vacancy state deeply modifies the conduction
properties  of the constriction. Indeed  Fig.~\ref{quantumdot} shows
at the Dirac energy a well defined peak of conductance
corresponding to the vacancy state.

When the defect is applied within the short nanoribbon, like in the
positions indicated as a), c) or d) in Fig.~\ref{quantumdot}, the
transmission  peak reaches the full value 1. The total 
conductance characteristics is also affected by the lateral
position of the defect. A defect whose position is shifted toward the
contact, as in case b) of Fig.~\ref{quantumdot}, produces the
conductance characteristics of the simple constriction plus a defect
peak at the Dirac energy  which does not reach the full quantum of
conductance. The heigth of the peak lowers when shifting the defect
more distant from the nanoribbon within the graphene lead.

The fact that the
contacts work as diffraction barriers decouples the central device
from the leads, and hence the vacancy state from the leads
states. The transmission by the vacancy mode through a
double barrier can be modeled by  the
standard Lorentzian approximation \cite{G}:

     \begin{equation}
     T(E)= \frac{4\Gamma_{1}\Gamma_{2}}{\Gamma^{2}}
\left[\frac{\Gamma^{2}}{(E-E_{R})^{2} +\Gamma^{2}} \right]
\label{TT}
\end{equation}
where $E_{R}$ is the resonance energy and $\Gamma=(\Gamma_{1} +
\Gamma_{2})/2$, where $\Gamma_{1}$ and $\Gamma_{2}$ measure the
coupling of the resonant state to the leads.  For a symmetric 
position of the functionalized atom
with respect to the left and right contacts, like in the positions
indicated as a), c) or d) in Fig.~\ref{quantumdot}, the double
barrier is symmetric. We have
$\Gamma_{1}=\Gamma_{2}=\Gamma$ and the maximum transmission
$\Gamma_{1}\Gamma_{2} / \Gamma^{2}$ is one. For an  asymmetric position of the
functionalized atom, like in the position b) in
Fig.~\ref{quantumdot}, the double
barrier is asymmetric and the maximum transmission is smaller than
one. The maximum transmission in the peak tends to zero when the ratio
$\Gamma_{1}/\Gamma_{2}$ tends to zero or to infinity, {\it i.e.} when
shifting the defect more and more within the graphene lead.

Let us emphasize also that in a metallic nanotube functionalized by
an OH group, a state (for example a $\pi-\pi^*$ bond) is removed from
a conduction level and the conductance becomes 0 at
the level of the removed state \cite{H}. Instead, when
functionalizing within diffraction barriers like in the present case,
the conductance is enhanced rather than destroyed.

  Let's now discuss
experimental implications of our findings on quantum dots
and suggest new experiments as well as possible technological devices.

We find that the computed
characteristics of the geometric quantum dot (Fig. \ref{quantumdot1})
looks very similar to what has
been effectively measured in a recent experiment done on a graphene
quantum dot (see Ref.~\cite{Geim}).  The present 
work gives important insight  in  the role of difffraction to produce 
the electron confinement in the dot, in particular for energies close
to the Dirac point.

More generally  we note that
the control of the characteristics of existing  quantum dot is always  a
difficult task  and the graphene quantum dot discussed here offers new
possibilities. For instance, using appropriate electric gates
\cite{Ensslin} it is  possible to adapt the wavelength of the
incident and outcoming electrons in the half planes. This will
modulate  the characteristics of the barriers that couple the quantum
dot to the leads. It is then possible to go from a small to a large
barrier transparency, and thus to study a transition from a
localized to a metallic regime.

Finally the functionalized graphene constriction could  have
applications in spintronics. Indeed, with a spin polarized defect,   the quantum dot  allows the production of spin 
polarized currents. Control over the spin of the defect by a magnetic 
field could provide
control over the spin current providing a molecular spin
valve.


To conclude our  work establishes an important  link between 
nanoelectronics and subwavelength optics, that could be 
experimentally tested thanks to the remarquable properties of 
graphene.  The concept of "electronic diffraction barrier" is central 
to the understanding of transport properties of subwavelength 
graphene nanodevices. It allowed us in particular to get insigth in 
the properties of quantum dots and to propose a new kind of 
technological device for molecular spin valves. This shows the high 
potential of the bridging between  subwavelength optics and 
nanoelectronics with graphene as a constituent material.  

\textit{Acknowledgements} -
This work has benefited from exchanges with many colleagues. We thank
in particular  F.~Balestro, X.~Blase, D.~Feinberg, J.~Le~Perchec,
L.~L{\'e}vy, L.~Magaud, P.~Mallet,  C.~Naud, T.~Lopez-Rios,  P.~Qu{\'e}merais,
F.~Varchon, J.~Y.~Veuillen and W.~Wernsdorfer. We also thank C.~
Berger \& W.~de~Heer for a critical reading of the manuscript. Computer time has 
been granted by
IDRIS and CIMENT/PHYNUM.


\begin{thebibliography}{99}


 \bibitem{G}
Datta, S.
{\it Electronic Transport in Mesoscopic Systems}
(Cambridge University Press, Cambridge, 1995).

\bibitem{Wolf}
Wolf, P.-E. \&  Maret, G.
Weak localization and coherent backscattering of photons in disordered media.
{\it Phys. Rev. Lett.} {\bf 55,} 2696 (1985).


\bibitem{Altshuler}
Cheianov, V.~V., Fal'ko, V. \&  Altshuler, B.~L.
The Focusing of Electron Flow and a Veselago Lens in Graphene p-n Junctions.
\textit{Science} \textbf{315,} 1252 (2007).


\bibitem{Novoselov}
Novoselov, K.~S. et al.
Two-dimensional gas of massless Dirac fermions in graphene.
{\it Nature} {\bf 438,} 197-200 (2005).

\bibitem{Kim}
Zhang, Y., Tan, Y.-W., Stormer, H.~L. \& Kim, P.
Experimental observation of the quantum Hall effect and Berry's phase in
graphene.
{\it Nature} {\bf 438,} 201-204 (2005).

\bibitem{Berger}
Berger, C. et al.
Electronic confinement and coherence in patterned epitaxial graphene.
{\it Science} {\bf 312,} 1191-1196 (2006).

\bibitem{Berger2}
de~Heer, W.~A. et al.
Epitaxial graphene.
{\it Solid State Com.} {\bf 143,} 92-100 (2007).

\bibitem{A}
Bolotin, K.~I. et al.
Ultrahigh electron mobility in suspended graphene.
{\it Solid State Com.} {\bf 146,} 351-355 (2008).

\bibitem{B}
\"Ozyilmaz, B. et al.
Electronic transport and quantum Hall effect in bipolar graphene p-n-p
Junctions.
{\it Phys. Rev. Lett.} {\bf 99,} 166804 (2007).

\bibitem{B2}
Li, X.,  Wang, X., Zhang, L., Lee, S. \&  Dai, H.
Chemically Derived, Ultrasmooth Graphene Nanoribbon Semiconductors.
{\it Science} {\bf 319,} 1229 (2008).


\bibitem{Geim}
Ponomarenko, L.~A. et al.
Chaotic Dirac billiard in graphene quantum dots.
{\it Science} {\bf 320,} 356 (2008).


 \bibitem{Ozbay}
 Ozbay, E.
 Plasmonics: Merging Photonics and Electronics at nanoscale Dimensions.
 {\it Science} {\bf 311,} 189 - 193 (2006).
 
 \bibitem{Barnes}
Barnes, W.~L.,Dereux, A. \& Ebbesen, T.~W. 
Surface plasmon subwavelength optics.
{\it Nature} {\bf 424,} 824 - 830 (2003).

\bibitem{C}
Wallace, P.~R.
The band theory of graphite.
{\it Phys. Rev.} {\bf 71,} 622 (1947).


 \bibitem{Ebbesen}
Ebbesen, T.~W.,Lezec, H.~J., Ghaemi, H.~F., , Thio, T., \& Wolf, P.~E. 
 Extraordinary optical transmission through sub-wavelength hole arrays.
{\it Nature} {\bf 391,} 667 - 669 (1998).

\bibitem{D}
Geim, A.~K. \& Novoselov, K.~S.
The rise of graphene.
{\it Nature Materials} {\bf 6,} 183 - 191 (2007).

\bibitem{Bethe}
Bethe, H.~A.
Theory of diffraction by small holes.
{\it Phys. Rev.} {\bf 66,} 163 (1944).


\bibitem{Footnote2}
Peres, N.~M.~R., Castro Neto, A.~H. \& Guinea, F.
Dirac fermion confinement in graphene.
{\it Phys.~Rev.~B} {\bf 73,} 241403(R) (2006).


\bibitem{Cavity2}
Qu\'emerais, P., Barbara, A., Le~Perchec, J. \& L\'opez-Ríos, T.
Efficient excitation of subwavelength metallic gratings.
\textit{J. App. Phys.} \textbf{97,} 053507 (2005).

\bibitem{Takakura}
Takakura,Y.
Optical Resonance in a Narrow Slit in a Thick Metallic Screen.
{\it Phys.~Rev.~Lett.} {\bf 86,} 5601-5603 (2001).

\bibitem{Suckling}
Suckling,J.~R et al.
Finite Conductance Governs the Resonance Transmission of Thin Metals
Slits at Microwave Frequencies.
{\it Phys.~Rev.~Lett.} {\bf 92,} 147401(2004).




\bibitem{dresselhaus}
Nakada, K., Fujita, M., Dresselhaus, G. \& Dresselhaus, M.~S.
Edge state in graphene ribbons: Nanometer size effect and edge shape
dependence.
{\it Phys.~Rev.~B} {\bf 54,} 17954 (1996).


\bibitem{J}
Lin, Y.-M., Perebeinos, V.~, Chen, Z.~ \& Avouris,  P.~
Conductance quantization in graphene nanoribbons.
arXiv:0805.0035.

\bibitem{K1}
     Mucciolo, E.~R., Castro Neto, A.~H. \& Lewenkopf, C.~H.
Conductance quantization and transport gap in disordered graphene
nanoribbons.
arXiv:0806.3777.

\bibitem{K2}
Lherbier, A., Biel, B.,  Niquet, Y.-M., \& Roche, S.
Transport length scales in disordered graphene-Based materials:
Strong localization regimes and dimensionality effects.
{\it Phys. Rev. Lett.} {\bf 100,} 036803 (2008).





\bibitem{E}
Lee, Y.-S. \& Marzari, N.
Cycloaddition functionalizations to preserve or control the
conductance of carbon nanotubes.
{\it Phys. Rev. Lett.} {\bf 97,} 116801 (2006).

\bibitem{F}
Pereira, V.~M., Guinea, F., Lopes dos Santos, J.~M.~B., Peres,
N.~M.~R.,  \&  Castro Neto, A.~H.
Disorder Induced Localized States in Graphene.
{\it Phys. Rev. Lett.} {\bf 96,} 036801 (2006).
%

\bibitem{H}
Choi, H.~J., Ihm, J., Louie, S.~G. \&  Cohen, M.~L.
Defects, quasibound states, and quantum conductance in metallic
carbon nanotubes.
{\it Phys. Rev. Lett.} {\bf 84,} 2917 (2000).


\bibitem{Ensslin}
Stampfer, C. et al.
Tunable Coulomb blockade in nanostructured graphene.
{\it Appl. Phys. Lett.} {\bf 92,} 012102 (2008).

\end{thebibliography}
\end{document}